\documentclass[prl]{revtex4}

\begin{document}

\title{Polarization modulation instability in liquid crystals with
spontaneous chiral symmetry breaking }
\author{Nata\v{s}a Vaupoti\v{c}$^{1,3}$ and Martin \v{C}opi\v{c}$^{2,3}$}
\affiliation{$^{1}$Department of Physics, Faculty of Education, University of
Maribor,Maribor, Slovenia\\
$^{2}$Department of Physics, Faculty of Mathematics and Physics, University
of Ljubljana, Ljubljana, Slovenia\\
$^{3}$Jozef Stefan Institute, Ljubljana, Slovenia}

\begin{abstract}
We present a theoretical model which describes the polarization modulated
and layer undulated structure of the B7 phase and gives the phase transition
from the synclinic ferroelectric B2 phase to the B7 phase as observed
experimentally. The system is driven into the modulated phase due to the
coupling between the polarization splay and the tilt of the molecules with
respect to the smectic layer normal. The modulation wavelength and the width
of the wall between two domains of opposite chirality are estimated.
\end{abstract}

\maketitle

The discovery of the polar order and macroscopic chirality in smectic liquid
crystals of bow-shaped (also called bent-core or banana-shaped) molecules is
one of the most fascinating features found in liquid crystals in the last
decade. These systems represent the first example of the formation of chiral
structures without possessing chirality on the molecular level \cite%
{Niori,Link-Science}. The polar order and the macroscopic chirality appear
spontaneously as a result of broken orientational symmetries. The most
widely studied phase formed by bow-shaped molecules is the B2 phase which is
layered and the molecules inside the layer are tilted with respect to the
layer normal. This tilt, together with polarization and the layer normal,
breaks the chiral symmetry. The tilt in the neighboring layers can be either
in the same (synclinic) or in the opposite (anticlinic) direction. Close
packing of bow-shaped molecules results in polar order of each smectic
layer. The neighboring layers can be either synpolar (ferroelectric) or
antipolar (antiferroelectric). So, in all, four different structures are
possible. The adopted nomenclature for these B2 phases is smectic C$_{S,A}$P$%
_{F,A}$, where the indices S and A behind the C stand for the synclinic or
anticlinic tilt in the adjacent layers and F and A next to P stand for
ferroelectric or antiferroelectric ordering of polarization in the adjacent
layers.

Since the molecules are achiral the tilt of the molecules does not affect
the polarization significantly and the latter is determined primarily by the
magnitude of the dipole moment of the constituent molecules. Recent
experimental findings that polar order can exist also without the tilt \cite%
{Eremin} suggest that polar ordering is a fundamental property of liquid
crystals formed by bow-shaped molecules while the tilt of the molecules with
respect to the layer normal depends on the particular molecular structure.

Theoretical models of the bow-shaped molecular systems are scarce and the
origin of the molecular tilt and the layer polarity is still not clear. The
polar ordering is most probably directly driven by the strongly polar
molecular shape \cite{Osipov}, and spontaneous polarization occurs due to
the polar excluded volume effects \cite{Lansac}. An exhaustive
classification of the symmetry-allowed smectic phases was presented in \cite%
{Brand}, a phenomenological Landau model that produces many of the banana
phases was introduced in \cite{Roy}. In \cite{Lubensky} it was shown that
three order parameters (the first, the second and the third-rank tensor) are
necessary to fully characterize the phases exhibited by bow-shaped molecules.

Recently, the B7 phase, which shows extremely rich and fascinating textures
and has a more complicated structure than the B2 phase, received
considerable attention. J\'{a}kli \textit{et al}. \cite{Jakli-prl} suggested
that the structure of the B7 phase is identical to the Sm-Cg phase, which is
lamellar with triclinic local layer symmetry. The Boulder group suggested a
different structure of the B7 phase: it is a polarization splayed and layer
modulated structure \cite{Clark-Science} and their studies of the electric
field induced transition between the polarization modulated and the
ferroelectric smectic C$_{S}$P$_{F}$ phase gave no evidence of the Sm-Cg
ordering.

In this letter we present a theoretical model which shows the existance of an instability in the 
Sm-C$_{S}$P$_{F}$ phase that leads to a spatially modulated phase with splayed polarization just
as observed experimentally \cite{Clark-Science}.

Since the polarization modulated structure has been observed only in
ferroelectric systems where the tilt in the adjacent layers is synclinic and
since the preferred local orientation of the polarization is perpendicular
to the tilt plane, we propose that the B7 phase and the Sm-C$_{S}$P$_{F}$
phase can be described by an extension of the model that was developed by
the present authors to describe the structure in confined ferroelectric
liquid crystals \cite{pre96,pre2003}. Within this model the smectic
structure is described in terms of the smectic order parameter $\psi $, the
nematic director $\mathbf{n}$, which describes the average local orientation
of the axes that go through the top and the bottom of the molecules (see
Fig. \ref{fig1}) and the polar parameter $\mathbf{p}$, which is the unit
vector in the direction of the local polarization $\mathbf{P}$. Due to the
bow-shaped structure of the constituent molecules, the dipole moment of the
molecule is always perpendicular to the molecular long axis. Because of that
we assume that the angle between the nematic director and the dipole moment
of a cluster of molecules described by the nematic director is also fixed,
and we set $\mathbf{n}\cdot \mathbf{p}=0$.

Due to the vectorial symmetry of the polarization the local free energy
contains terms of the form $(\nabla \cdot \mathbf{P})^{i}$, where $i$ is an
integer. This terms can stabilize a finite splay of polarization. Since the
constituent molecules have thicker cores than tails, one type of splay
(positive or negative) is privileged. The use of the terms with $i=1$ and $%
i=2$ has already been proposed \cite{Clark-Science}. However, the linear
term $\nabla \cdot \mathbf{P}$ is a surface term, and the quadratic term,
even if its coefficient is negative, cannot stabilize the polarization
modulation in bulk sample, such that the volume average $<\nabla \cdot 
\mathbf{P>}\neq 0$. So we write the free energy density in the following
form:

\begin{eqnarray}
f &=&\frac{1}{2}K_{n}\left[ (\nabla \cdot \mathbf{n)}^{2}+\left( \nabla
\times \mathbf{n}\right) ^{2}\right] +c_{||}|(\mathbf{n}\cdot \nabla
-iq_{0})\psi |^{2}+c_{\bot }|\mathbf{n}\times \nabla \psi |^{2}+D|\left( 
\mathbf{n}\times \nabla \right) ^{2}\psi |^{2}+ \nonumber \\
&&\frac{P_{0}^{2}}{2\varepsilon
\varepsilon _{0}}\ p_{x}^{2}+\widetilde{K}_{p}(\nabla \cdot \mathbf{p)}|\mathbf{n}\times \nabla \psi
|^{2}\mathbf{+}\frac{1}{2}K_{p}(\nabla \cdot \mathbf{p)}^{2}+K_{np}|\mathbf{p%
}\times (\mathbf{n}\times \nabla \psi )|^{2}\ .
\label{enf}
\end{eqnarray}%
The first four terms are the terms that are used to describe the structure
in ordinary, achiral, smectic liquid crystals. The first term gives the
nematic elastic energy density. The second term describes the
compressibility of the smectic layers and defines the smectic layer
thickness. The parameter $q_{0}$ is defined as $q_{0}=2\pi /d_{0}$, where $%
d_{0\text{ }}$is the smectic layer thickness in the nontilted phase. In the
tilted phases $c_{\bot }$ is negative and the $D$ term stabilizes a finite
tilt of $\mathbf{n}$ with respect to the layer normal $\mathbf{\nu }$. The
fifth term presents the energy due to the dipole-dipole self-interaction,
when all the variables are function of the $x$-coordinate only. It is
important in chiral smectics and bow-shaped systems with large value of the
spontaneous polarization. $P_{0}$ is the magnitude of the local
polarization, $\varepsilon $ is the dielectric constant and $p_{x}$ is the $%
x $-component of the polar parameter.

The last three terms in Eq. (\ref{enf}) are new and describe the
characteristics of the phases formed by bow-shaped molecules. The term with $%
\widetilde{K}_{p}$ is the lowest order coupling term between the splay of
polarization and the tilt of the director with respect to the layer normal.
If the coupling is large enough, polarization modulated phase becomes stable
with respect to the homogeneous (unmodulated) phase. The coupling term prefers negative
splay if the parameter $\widetilde{K}_{p}$ is positive. The term with $K_{p}$
stabilizes a finite splay of polarization. The last term is the coupling
term between the polar parameter, the layer normal and the nematic director.
This term prefers the direction of polarization being perpendicular to the
plane defined by the layer normal and the director $\mathbf{n}$ (in ordinary
chiral SmC polarization is always perpendicular to this plane).

The smectic order parameter is expressed as $\psi =\eta \exp
\{iq_{t}(z+u(x)\}$, where $q_{t}$ is the smectic layer periodicity along the 
$z$-axis, $u(x)$ is the layer displacement from the bookshelf geometry of
layers and $\eta $ is the magnitude of the smectic order parameter and we
assume that it is constant. The layer displacement is related to the tilt $%
\Delta $ of the smectic layers as $du(x)/dx=-\tan \Delta $. We express the
free energy density (\ref{enf}) in terms of the angles $\alpha $, $\varphi $%
, $\vartheta $ and $\Delta $ (see Fig. \ref{fig1}) and expand it around the
homogeneous structure with bookshelf geometry of smectic layers ($\vartheta
=\vartheta _{B}$, $\varphi =\Delta =\alpha =0$) up to the second order terms
in $\delta \vartheta ,\delta \varphi $, $\delta \Delta $ and $\delta \alpha $%
. In order to check under which conditions the homogeneous structure becomes
unstable we study the effect of the fluctuations with a wave vector $q$ on
the structure. The variables are expanded as $\delta V=\sum_{q}V_{q}\exp
(iqx)$ where $V$ stands for any of the four variables $\vartheta $, $\alpha $%
, $\varphi $ and $\Delta $. The quadratic part of the variation in the free
energy $F$ is finally expressed as: 
\[
\delta ^{2}F=\sum_{q>0}\left( 
\begin{tabular}{l}
$\alpha _{-q}$ \\ 
$\varphi _{-q}$ \\ 
$\vartheta _{-q}$ \\ 
$\Delta _{-q}$%
\end{tabular}%
\ \right) \underline{M}_{2}\left( 
\begin{tabular}{l}
$\alpha _{q}$ \\ 
$\varphi _{q}$ \\ 
$\vartheta _{q}$ \\ 
$\Delta _{q}$%
\end{tabular}%
\ \right) \ . 
\]%
If the matrix \underline{$M$}$_{2}$ has only positive eigenvalues the
homogeneous structure is stable. Our main finding is that at some $q_{cr}$
one of the eigenvalues becomes negative, so a modulated structure appears.
The critical wave vector $q_{cr}$ can be expressed analytically, however it
is a very complicated function that depends on all the parameters in the
free energy density. In general the relations are the following:

\begin{itemize}
\item The homogeneous structure becomes unstable if (all the rest of the
parameters being fixed) the value of $\widetilde{K}_{p}$ is greater than the
critical value $\widetilde{K}_{p}^{cr}$. In Fig. \ref{fig2} we show
analytically obtained results for the dependence of $\widetilde{K}_{p}^{cr}$
and $q_{cr}$ (at $\widetilde{K}_{p}=\widetilde{K}_{p}^{cr}$) on $K_{p}$ at
different values of $\vartheta _{B}$. The value of $\widetilde{K}_{p}^{cr}$
decreases if $\vartheta _{B}$ increases or/and if $K_{p}$ decreases.

\item The eigenvector that corresponds to the critical value of $q$ at the
critical value of $\widetilde{K}_{p}$ has the components of deformations in $%
\vartheta $ and $\Delta $ much larger than the components of deformations in 
$\varphi $ and $\alpha $. As expected the magnitude of deformation in $%
\alpha $ increases if $K_{np}$ decreases. The deformation in $\varphi $ is
negligibly small.

\item Modulated structure results due to the coupling between the
polarization splay and the director tilt $\vartheta $. There are other
couplings possible (e.g. between the splay of polarization and the layer
compressibility), however they all give higher order instabilities. The same
is true for the term $(\nabla \cdot \mathbf{P})^{3}$. Therefore we conclude
that the tilt is essential to obtain polarization modulated structure.
\end{itemize}

At $\widetilde{K}_{p}>\widetilde{K}_{p}^{cr}$ the symmetry-required local
preference for polarization is to be splayed. However it is impossible to
achieve splay of the preferred sign everywhere in space unless appropriate
walls are introduced (Fig. \ref{fig3}). Two types of walls are expected,
depending on whether the chirality switches across the wall or not. 

Chirality switching occurs when the molecules rotate around the long axis.
Due to the bow shape of the molecules this rotation is strongly hindered
when the molecules are tilted. In order to rotate around the long axis the
cone angle has to decrease significantly and we assume that it goes to
zero. Spatial variation in the cone angle is coupled to the layer
deformation. The situation across the wall can, in the most simplified
version, be described by no spatial variation of the director and the
polarization director across the wall (see Fig. \ref{fig3}). Then the free
energy density has only three spatially dependent terms: 
\[
f=-|c_{\bot }|q_{0}^{2}\eta w^{2}+Dq_{0}^{4}\eta ^{2}w^{4}+Dq_{0}^{2}\eta
^{2}\left( \frac{dw}{dx}\right) ^{2}\ ,
\]%
where $w=du/dx$. The Euler-Lagrange equation that follows from the
minimization of the free energy across the wall can be solved by the ansatz $%
w=w_{0}\tanh (x/\lambda _{w} )$, where $\lambda _{w}$ is the characteristic
width of the wall, and one finds 
\[
\lambda _{w}=\sqrt{\frac{2D}{|c_{\bot }|}}=\frac{1}{q_{0}\tan \vartheta _{B}}%
\ .
\]%
To estimate the characteristic widths the following set of parameters is
used: $P_{0}=300\ \mathrm{nC/cm}^{2}$, $\varepsilon =10$, $\vartheta
_{B}=40^{\circ }$, $|c_{\bot }|/c_{||}=0.1$ and $K_{n}=K_{p}\sim 10^{-11}\ 
\mathrm{N}$. We define the layer compressibility constant as $%
B=c_{||}q_{0}^{2}\eta ^{2}\sim 10^{4}\ \mathrm{Jm}^{-3}$. The
compressibility is rather low compared to the ordinary smectics \cite%
{Clark-Science}. The smectic penetration depth is then $\lambda _{||}=\sqrt{%
K_{n}/B}\sim 30\ \mathrm{nm}$. With the chosen set of parameters the width
of the wall over which chirality switches is $\lambda _{w}=0.2d_{0}$, i.e.
only a few widths of the constituent molecules. The width of the wall over
which chirality does not change is essentially of the same order of
magnitude, however the energy associated with the wall is lower since there
is no need for the cone angle to reduce to zero (see Fig. \ref{fig3}).

Finally we estimate the modulation length ($\lambda _{m}$), i.e. the width
of the stripe with the preferred polarization splay. In a very crude
estimate we consider only the terms that contain spatial derivative of the $x
$-component of the polarization vector and set $\vartheta $ to equal its
bulk value. We find that the equilibrium value is $dp_{x}/dx\sim \widetilde{K%
}_{p}$\smallskip $q_{0}^{2}\eta ^{2}\sin ^{2}\vartheta _{B}/K_{p}$. Setting $%
dp_{x}/dx=1/\lambda _{m}$ and using the critical value of $\widetilde{K}_{p}$
at $K_{p}/K_{n}=1$ (see Fig. \ref{fig2}b) we obtain $\lambda _{m}\approx
80\ \mathrm{nm}$. The modulation length can also be estimated from the
critical value of $q$ at the critical value of $\widetilde{K}_{p}$ (see Fig. %
\ref{fig2}a). One finds $\lambda _{m}=2\pi /q_{cr} \approx
10\ \mathrm{nm}$. Both values agree qualitatively with the
experiment where the observed modulation lengths are of the order of 10
layer thicknesses.

To conclude we have presented a theoretical model which describes the phase
transition from the smectic-C$_{S}$P$_{F}$ phase to the layer undulated and
polarization modulated phase. The
system is driven into the modulated phase due to the coupling between the
polarization splay and the tilt of the molecules. As the tilt is responsible
for the breaking of chiral symmetry, the instability is due to the coupling
between polarization and chirality. The transition from the homogeneous to
the modulated phase occurs if the coupling term is strong enough. Since this
coupling depends on the structure of the constituent molecules, we predict
that a small change in the molecular structure, e.g. in the molecular tail
can lead to a significant change in the coupling and thus determine whether
the system is in the homogeneous or the modulated phase.

\begin{figure}
\caption{The local arrangement of the constituent
molecules is described by the nematic director $\mathbf{n}$, by the polar
parameter $\mathbf{p}$ and the smectic layer normal $\mathbf{\protect\nu }$.
The orientation of these three vectors is described by the cone angle $%
\protect\vartheta $ (the tilt of the director with respect to the smectic
layer normal), the smectic layer tilt $\Delta $, the director position on
the cone $\protect\varphi $ and the angle $\protect\alpha $ which describes
the tilt of polarization from the direction perpendicular to the tilt plane
(the plane determined by the director $\mathbf{n}$ and the smectic layer
normal $\mathbf{\protect\nu }$).}
\label{fig1}
\end{figure}

\begin{figure}
\caption{a) The dependence of $q_{cr}$ and b) the dependence of the critical value of the
coupling constant $\widetilde{K}_{p}$ between the polarization splay and the
director tilt on the ratio between the
polar and nematic elastic constant. Solid line: $\protect\vartheta %
_{B}=20^{\circ }$; dotted line: $\protect\vartheta _{B}=30^{\circ }$; dashed
line: $\protect\vartheta _{B}=40^{\circ }$. Parameter values: $|c_{\bot
}|/c_{||}=0.1$, $B/(q_{0}^{2}K_{n})=7\times 10^{-4}$, $P_{0}^{2}/(2\protect\varepsilon 
\protect\varepsilon _{0}B)=5$, $K_{np}q_{0}^{2}\protect\eta ^{2}/B=10$.}
\label{fig2}
\end{figure}

\begin{figure}
\caption{In a ferroelectric domain stripes of the
preferred polarization splay are divided by walls. The neighboring stripes
can have the same or the opposite chirality. Below is shown the structure of
the wall over which chirality switches and above the structure of the wall
with no chirality switch is shown. Figure presents the top view on the
undulated layer, and at the wall (of width $\protect\lambda _{w}$) there is
the top of the hill or the bottom of the valley. The radius of the circle is
a measure for the cone angle, the arrows present the polar parameter and the
lines with bars present the director position on the cone.}
\label{fig3}
\end{figure}

\end{document}